\documentclass[aps,floats, amsmath,epsf,onecolumn, prl, superscriptaddress]{revtex4}
\usepackage{calc}
\usepackage{psfrag}
\usepackage{graphicx}
\usepackage{color}
\usepackage{bbold}
\usepackage{bm}
\usepackage{slashed}
\begin{document}

\title{Note on ``Quantum superconducting criticality in graphene and
topological insulators"}

\author{Bitan Roy}
\affiliation{Condensed Matter Theory Center and Joint Quantum Institute, University of Maryland, College Park, Maryland 20742-4111, USA}

\author{Vladimir Juri\v ci\' c}
\affiliation{Nordita,  Center for Quantum Materials,  KTH Royal Institute of Technology and Stockholm University, Roslagstullsbacken 23,  10691 Stockholm,  Sweden}

\author{Igor F. Herbut}
\affiliation{ Department of Physics, Simon Fraser University, Burnaby, British Columbia, Canada V5A 1S6}

\begin{abstract}
We correct our previous conclusion regarding the fate of a charged quantum critical point across the superconducting transition for two dimensional massless Dirac fermion. Within the leading order $\epsilon$ expansion, we now find that the requisite number of four-component Dirac fermion flavors ($N_f$) for the continuous phase transition through a charged critical point is $N_f>18.2699$. For $N_f\geq1/2$,  the critical number of bosonic flavors for this transition is significantly reduced as compared to the value determined in the absence of the Dirac fermions in the theory.
\end{abstract}
\maketitle

We here take the opportunity to correct an error in the expression from Eq.~(18) of the ``Supplementary Materials" (SM) of Ref.~\cite{RJH}. The correct expression should read as
\begin{eqnarray}
(4a)=- (-g) (i e)^2 \; \int \frac{d^d q}{(2 \pi)^d} \: \frac{1}{ q^2} \: \left( \delta_{\mu \nu} - \frac{q_\mu q_\nu}{q^2}\right) \gamma_5 \gamma_\mu
\frac{\slashed{k}-\slashed{q}}{(k-q)^2} \: P_+ \: \frac{\slashed{k}+\slashed{k}'-\slashed{q}}{(k+k'-q)^2} \; \gamma_5 \gamma_\nu,
\end{eqnarray}
where $\slashed{k}=k_\mu \gamma_\mu$. The evaluation of this expression has already been shown in the SM (notice the extra minus sign in the above expression) of the original article, which then leads to a slightly modified renormalization condition for the Yukawa coupling in Eq.~(10) of the main part of the paper
\begin{equation}\label{eq:renorm-condition-yukawa}
Z_\Psi Z_\Phi^{1/2} g_0 \mu^{-\epsilon/2}-3e^2g\frac{1}{\epsilon}=g.
\end{equation}
Notice that instead of the expression as given in the paper, the above equation contains a negative sign in the term $\sim e^2 g$.
The renormalization factors $Z_\Psi$ and $Z_\Phi$ are given by Eq.~(8) in the original paper. As a result, the correct form
of the $\beta$-function for the Yukawa coupling is
\begin{equation}
\beta_{g^2}=-\epsilon g^2 +(N_f+1)g^4 - 6e^2 g^2,
\end{equation}
which differs in the value of the coefficient of the last term as compared to Eq.~(14) of the original article and is in agreement with Ref.\ [\onlinecite{Maciejko}] when we set $N_f=1/2$ (as the authors of Ref.~\cite{Maciejko} addressed superconducting criticality for the surface states of topological insulators). Rest of the flow equations remains unchanged and together they support a pair of charged fixed points located at
\begin{equation}
\left( e^2_\ast, g^2_\ast, \lambda^\pm_\ast \right)= \left( \frac{3}{4 Y}, \frac{9+2Y}{2 X Y}, \frac{\Delta_1 \pm \sqrt{\Delta^2_1+\Delta_2}}{2 X Y W} \right),
\end{equation}
analogous to Eq.~(19) of the original paper, where $X=N_f+1$, $Y=N_f+N_b$, $W=N_b+4$ and
\begin{equation}
\Delta_1= XY+18 X-N_f(9+2 Y), \quad \Delta_2=-4 W \left[ 54 X^2-2 N_f (4.5+Y)^2 \right].
\end{equation}
Note that expressions for $\Delta_1$ and $\Delta_2$ is now slightly modified due to the change in renormalization condition in Eq.~(\ref{eq:renorm-condition-yukawa}).

\begin{figure}[htb]
\includegraphics[width=6.5cm,height=5.0cm]{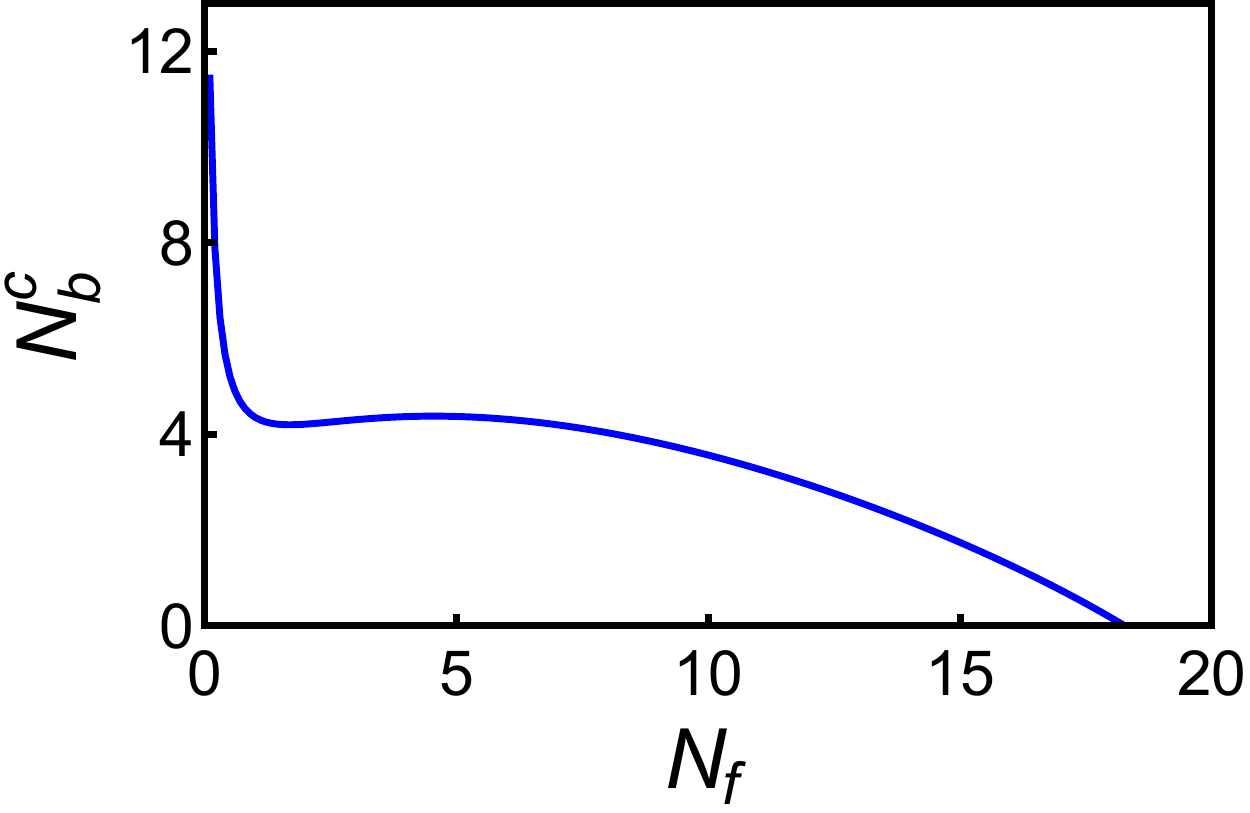}
\caption{Critical number of bosonic flavors ($N^{c}_b$) for the continuous quantum phase transition through a
charged fixed point as a function of the number of four-component fermionic flavors ($N_f$). For $N_b>N_b^c$ the transition is continuous, while otherwise it is of the first order in nature (see the blue curve). When $N_f>18.2699$, the quantum phase transition through the charged fixed point is continuous for any number of bosonic flavors.}\label{fig_1}
\end{figure}

We thus obtain a different function determining the critical number of the bosonic flavors ($N^{c}_b$) as a function of the number of four-component fermionic flavors ($N_f$) for the second order quantum phase transition through a charged fixed point, as shown in Fig.~\ref{fig_1}. In particular, the quantum-critical point in the supersymmetric theory with $N_f=1/2$ and $N_b=1$ becomes unstable when both fermions and bosons are coupled to the gauge field. The transition then turns into the  first order one. However, for sufficiently large number of fermionic flavors the transition through the charged fixed point is continuous for arbitrary number of bosonic flavors ($N_b$).
\\

Due to this modification in Eq.~(\ref{eq:renorm-condition-yukawa}) the flow equation for the Ginzburg-Landau parameter $\kappa^2=\lambda/(2 e^2)$ from Eq.~(20) of the original article is changed accordingly. The flow equation of $\kappa^2$ gives $\kappa^2_+>0$ when $N_b> N_b^c$ for a given $N_f$.
\\

This result does not affect any other conclusions of the paper. However, few comments are due and they are stated below.

\begin{enumerate}

\item{Notice from Fig.~\ref{fig_1} that upon incorporating the coupling between fluctuating gauge fields and massless Dirac fermions, the critical number of bosonic flavors ($N^c_b$) for continuous phase transition through a charged critical point reduces \emph{drastically} even for sufficiently small number of fermionic flavors $N_f$. In particular, when $N_f \geq 0.5$ the critical number of bosonic flavors  drops to a value of $N^c_b\simeq 4-5$ from $N^c_b\sim 183$ determined in the absence of massless Dirac fermions in the theory~\cite{herbut-book}. Thus coupling of Dirac fermions with the gauge field increases the propensity of a continuous phase transition through a charged critical point.  }

\item{Since only one bosonic field couples with the fluctuating gauge field, it is worth determining the requisite fermionic flavor number for which $N^c_b=1$. From the leading order $\epsilon$-expansion we find $N_f \approx 16.55$ for $N^c_b=1$. }

\item{Although $N_f \approx 16.55$ seems too large to realize the existence of a charged critical point in any physical systems, it should be noted that upon incorporating next to leading order corrections ($\sim \epsilon^2$, for example) in the flow equations the above mentioned value of $N_f$ can change quite dramatically. Such analysis for our problem is not presently available. Nevertheless, it is worth pointing out that $N^c_b$ becomes \emph{negative} (suggesting continuous phase transition through charged critical point for any value of $N_b$) even in the absence of massless Dirac fermions, when corrections $\sim \epsilon^2$ are taken into account and $\epsilon$ is set to be \emph{unity} (for $d=3$) at the end of the anaysis~\cite{herbut-book}. Thus it remains as an open and interesting question to investigate the variation of critical fermion number for which a charged critical point can be accessed even for $N_b=1$, once the higher order corrections are taken into account.  }

\end{enumerate}

\end{document}